# Magnetic Properties of $RFe_2Si_2$ and $R(Fe_{1-x}M_x)_2Si_2$ Systems (R=La, Y and Lu, M= Ni, Mn and Cu

I. Felner[1], Bing Lv[2] and C.W. Chu[2]

[1]Racah Institute of Physics, The Hebrew University of Jerusalem, Jerusalem, 91904, Israel

[2]Department of Physics and the Texas Center for Superconductivity, University of Houston, Houston, Texas 77204-5005, USA

Abstract

*Due to the similarity to $BaFe_2As_2$ and $SrFe_2As_2$ the $RFe_2Si_2$ (R=La, Y and Lu) system has been proposed as a potential candidate for a new superconducting family containing Fe-Si layers as a structural unit. Various $R(Fe_{1-x}M_x)_2Si_2$ M=Ni, Mn and Cu) materials were synthesized and measured for their magnetic properties. None of these materials is superconducting down to 5 K. Fe in $RFe_2Si_2$ is paramagnetic. A pronounced peak at 232 K was observed in the magnetization curve of $YFe_2Si_2$. $^{57}Fe$ Mössbauer studies confirm the absence of any magnetic ordering at low temperatures. Similar peaks at various temperatures also appear in $R(Fe_{1-x}M_x)_2Si_2$ samples. Four independ factors affect the peak position and shift it to lower temperatures: (i) the lattice parameters, (ii) the concentration of x (iii) the applied magnetic field and (iv) the magnetic nature of M. The peak position is dramatically affected by the magnetic Mn dopants. It is propose that the magnetic peaks observed in $RFe_2Si_2$ and in $R(Fe_{1-x}M_x)_2Si_2$ represent a new nearly ferromagnetic Fermi liquid (NFFL) system and their nature is yet to be determined.*

PACS numbers: 74.10+v, 74.70.Wz, 75.50.Kj, 75.75.-c,

## (I) Introduction

Over the last four decades the ternary intermetallic compounds, which crystallize in the body-centered tetragonal (BCT) $ThCr_2Si_2$ (space group *I4/mmm*), have been of great interest due to the variety of physical phenomena observed in these compounds. As early as 1974, both magnetic and $^{57}Fe$ Mössbauer effect spectroscopy (MS) studies suggest that in $RFe_2M_2$ (R= La, Y and Lu and any other rare-earth element, M=Si or Ge), the Fe ions are not magnetically ordered [1]. Indeed, neutron powder diffraction measurements on $NdFe_2Si_2$ confirmed the absence of any magnetic moment on the Fe sites [2-3].

The recent discovery of superconductivity (SC) at relative high temperatures in F doped $SmFeAsO$ (1111), up to $T_C$= 55 K, [4] as well as in the doped $BaFe_2As_2$ and $SrFe_2As_2$ (122) system, has stimulated a large number of experimental and theoretical studies and intensified the search for high temperature superconductors (HTSC) in materials containing Fe-As layers as a structural unit. In the tetragonal $RFe_2Si_2$ (and in $BaFe_2As_2$ at elevated temperatures) (Fig.1) the R(Ba), Fe and Si(As) ions reside in the 2*a*, 4d, and 4*e* crystallographic positions, respectively.

The pristine $BaFe_2As_2$ sample, has a spin-density-wave (SDW) ground state at $T_N$=136(1) K. Band-structure calculations point to the fact that the SDW state arises on account of the special two-dimensional geometry of Fermi surface that is unstable to nesting [5]. The suppression of the SDW state by doping (either by electrons or holes), in most cases, induces SC in the system. Notably, both *electron* and *hole* doping are possible to generate SC and even by iso-valent atomic substitution, e.g., P for As, in stark contrast to the case of HTSC cuprates, in which only one type of carrier doping is possible to induce SC for one parent compound. Subsequently, partial substitution in $BaFe_2As_2$ of Ni or Co for Fe induces SC in the Ba(Fe$_{1-x}$Ni$_x$)$_2$As$_2$ and the Ba(Fe$_{1-x}$Co$_x$)$_2$As$_2$ systems [6-13]. A higher value for T$_C$= 38 K, was observed by optimal doping in the Ba$_{1-x}$K$_x$Fe$_2$As$_2$ system [6] and in pristine BaFe$_2$As$_2$ by application of high pressure [10]. Also associated with or preceding the magnetic transition is a tetragonal to orthorhombic structural transition, which is suppressed in the SC state. Perhaps, even more remarkable than the large T$_C$, is the large tenability these systems possess. It should be noted, that the suppression of the long-range magnetic ordering is **necessary** for the appearance of high-temperature SC, as in the layered Fe-As based systems and/or the cuprates HTSC. On the other hand, the major problem of the Fe-As based systems is the toxicity of As which limited their fabrication.

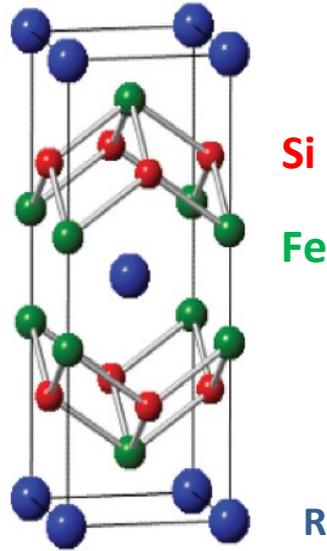

Fig. 1 (color online) The body-centered-tetragonal crystal structure of the RFe$_2$Si$_2$.

The above discoveries motivated a search for new superconductors of other systems having the ThCr$_2$Si$_2$-type structure. The close relation between the Fe-As and Fe-Si(Ge) systems, strongly suggest similar major role of Fe in the occurrence of SC. Therefore an extensive search for HTSC in the two similar RFe$_2$Si$_2$ and RFe$_2$Ge$_2$ systems is appealing. In both systems the Fe ions are not magnetic, thus no suppression of the SDW state mentioned above is needed. In addition both Si and Ge are cheap, non-toxic and very convenient materials.

The electronic structures of LaFe$_2$Si$_2$ and LaFe$_2$Ge$_2$ were calculated from first-principles. [14] Despite of the almost two-dimensionality of the crystal structure, the Fermi surface is three-dimensional. The density of states (DOS) at the Fermi level strongly depend on the distortion of the FeX$_4$ (X=Si and/or Ge) tetrahedra and/or the height of the Si (Ge) atom from the two-dimensional Fe plane. A steep increase of the Fe DOS are observed below the Fermi energy. It is claimed that this condition, creates excellent chances that doping in the Fe site will shift the DOS to the Fermi level, making the Fermi-surface two dimensional, condition which is essential for HTSC [14].

The desire to dramatically change or tune the properties of any compound, preferably through small changes in stoichiometry or composition, motivated us to search for new HTSC in doped RFe$_2$Si$_2$ (R=La, Y and Lu) systems in which R is a non-magnetic rare-earth element. Similar to SC induced in doped BaFe$_2$As$_2$ system, partial substitutions in RFe$_2$Si$_2$ in both R and Fe sites have been made. We have substituted tetravalent Ce for trivalent La, and Mn, Ni or Cu for Fe and measured their magnetic properties. All synthesized samples crystallize in the tetragonal ThCr$_2$Si$_2$ type structure. In this structure both Ni and Cu atoms are non-magnetic ions.[15] It appears that none of the compounds measured exhibits SC traces down to 5 K, although SC in Fe or Ni-containing 122 compounds was observed below 3 K, e.g. $YFe_{2-x}Si_2$ ($T_C$=3.0 K) [16]  $BaNi_2As_2$ ($T_C$=0.7 K) [17], $BaNi_2P_2$ ($T_C$=3.0 K) [18] and $LaNi_2P_2$ ($T_C$=1.8 K) [19].

In this paper we present the magnetic properties of $RFe_2Si_2$ systems in which doping was done in the Fe site only. The $La_{1-x}Ce_xFe_2Si_2$ system will be described elsewhere. Generally speaking, For R=Y and Lu, pronounced magnetic peaks appear at various temperatures.  For each peak, dc magnetic fields (H) as well as the dopant concentrations, affect dramatically its position and shift it to lower temperatures. For example: the sharp magnetic peak observed in $YFe_2Si_2$ at 232 K (measured at H=15 Oe), is shifted to 220 K when measured at H=1 kOe. This peak is shifted to 193 and 86 K in $Y(Fe_{1-x}Ni_x)_2Si_2$, for x=0.15 and x=0.5 respectively. In that sense, the magnetic behavior of $YNi_2Si_2$ (x=1) without a noticeable peak is reported here for the first time. On the other hand, in $Y(Fe_{1-x}Cu_x)_2Si_2$, the peak initially observed at T>300 K (for x=0.1), shifts back to lower temperatures for higher x values. Since the R atoms studied and Si are non-magnetic elements, the magnetic peaks are definitely related to the Fe sites. Our $^{57}Fe$ Mössbauer spectroscopy studies (MS) on various materials indicate clearly the absence of any permanent long-range ordering of Fe, even below the peaks position.  We propose that the peaks observed in $RFe_2Si_2$ and in their derivatives represent a new nearly ferromagnetic Fermi liquid (NFFL) systems, their nature will be discussed.

**(II) Experimental Details**

Polycrystalline samples with nominal composition $La(Fe_{1-x}Ni_x)_2Si_2$ (x=0, 0.5 and 0.15) $Y(Fe_{1-x}Ni_x)_2Si_2$ (x = 0, 0.05, 0.15 0.25, 0.50, 0.75 and 1), $Y(Fe_{1-x}Mn_x)_2Si_2$ (x=0.10, 0.20, 0.30, 0.5, 0.7 and 1), $Y(Fe_{1-x}Cu_x)_2Si_2$ (x=0.10, 0.25 and 0.5) and $Lu(Fe_{1-x}Ni_x)_2Si_2$ (x=0, 0.1, 0.2 and 0.3), were prepared by melting stoichiometric amounts of R, Fe and Si with Ni, Mn or Cu (all of at least 99.9% purity) in an arc furnace under high-purity Ar atmosphere. The arc-melted buttons were flipped and re-melted several times to ensure homogeneity. The samples, were structurally and chemically characterized by x-ray diffraction (XRD) using a Panalytical X'pert diffractometer, scanning electron microscopy and energy dispersive spectroscopy (EDS) using EDS-JOEL JSM-7700 SEM. The XRD powder patterns of all samples could be well indexed on the basis of BCT I4/mmm type structure. Magnetization measurements at various applied magnetic fields in the temperature interval 5 K < T < 300 K have been performed using the commercial (Quantum Design) superconducting quantum interference device (SQUID) magnetometer, with samples mounted in gel-caps. Prior to recording the zero-field-cooled (ZFC) curves, the SQUID magnetometer was always adjusted to be in a *"true"* H = 0 state. The temperature dependence of the field-cooled (FC) and the zero-field-cooled (ZFC) branches were taken via warming the samples from 5 K. The real ($\chi'$) and imaginary ($\chi''$) ac susceptibilities were measured with a home-made pickup coil method at a field amplitude of $h_0$=0.05 Oe at frequencies of $\omega/2\pi$ =56 and 1465 Hz. $^{57}Fe$ Mössbauer  spectroscopy (MS) studies at 90 and 295 (RT) were performed using a conventional constant acceleration drive, with a 50 mCi $^{57}Co$:Rh source. The spectra were analyzed in terms of least square fit procedures to theoretical expected spectra. The velocity calibration was performed with a α-iron foil at RT and the reported isomer shift (I.S.) values are relative to the Fe foil at RT.

**(III). Experimental results.**
 **(a) Structural and chemical characterization of $R(Fe_{1-x}Ni_x)_2Si_2$ and $Y(Fe_{1-x}Cu_x)_2Si_2$ samples.**

All XRD patterns obtained were indexed on the basis of a BCT structure with the lattice constants given in Table 1. A typical XRD pattern obtained for $YFe_2Si_2$ is shown in (Fig. 2) and Rietveld refinement on a long term XRD run yields z=0.371(1) for the free parameter of Si at the 4(e) site. The lattice parameters for YFe$_2$Si$_2$ are in fair agreement with $a$=3.910 and $c$=9.92 Å published in Ref 20. EDS chemical analysis of YFe$_2$Si$_2$ powder shows that the Y:Fe:Si ratio is exactly 1:2:2.

The R(Fe$_{1-x}$Ni$_x$)$_2$Si$_2$ system was extensively investigated and the XRD patterns indicate single phase materials which crystallize in the tetragonal structure. For small x values the $a$ lattice is practically constant, whereas the regular decrease in the $c$ lattice constant can be attributed to the smaller atomic radii of Ni (1.62 Å) as compared to that of Fe (1.72 Å). That leads to the decrease of unit cell volume with x (Table 1) and to the decrease of c/a ratio in Y(Fe$_{1-x}$Ni$_x$)$_2$Si$_2$ as shown in Fig. 2 (inset).

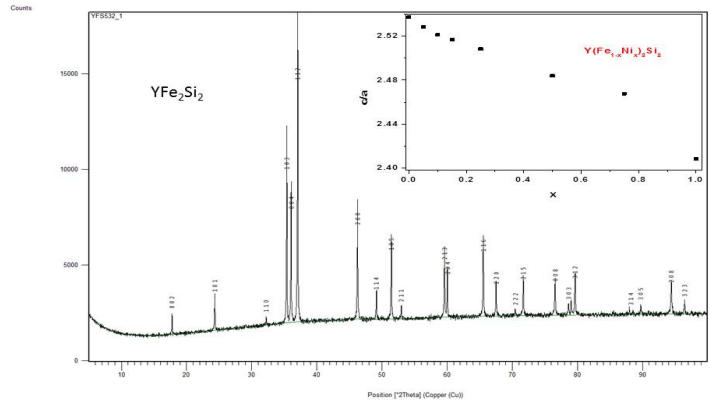

Fig. 2 XRD pattern of YFe$_2$Si$_2$. The inset shows the $c/a$ ratio for the Y(Fe$_{1-x}$Ni$_x$)$_2$Si$_2$ system.

Within the instrumental error, EDS studies of Y(Fe$_{1-x}$Ni$_x$)$_2$Si$_2$ confirm the nominal compositions of the Ni-doped samples; e.g. for x= 0.05, and 0.15 the measured Ni concentrations are: 0.06 and 0.17 respectively. On the other hand, Table 1 shows that the unit cell volume for Y(Fe$_{1-x}$Mn$_x$)$_2$Si$_2$ and Y(Fe$_{1-x}$Cu$_x$)$_2$Si$_2$ increases with x. For Mn (atomic radius 1.79Å) this increase is quite obvious, whereas for Cu (atomic radius 1.57Å) this increase needs more consideration and will be discussed later. In YFeCuSi$_2$ (x=0.5), additional extra lines in the XRD pattern which account to 5-7% of the spectral area, have been observed. These lines can be indexed on a basis of a hexagonal structure (AlB$_2$ type) which belong to YCuSi with a=4.031 and c=4.008 Å [21]. Extra lines were which account to 15% of the spectral area were observed in YFeMnSi$_2$ (x=0.5). They belong to an orthorhombic unit cell (S.G. Cmcm) with a= 4.047, b=15.83 and c=3.924 Å, similar to the structure of YFe$_{0.33}$Si$_2$.

*(b) Magnetic studies of* **RFe$_2$Si$_2$ (R= Y, La and Lu)**

Comprehensive magnetic measurements have been performed on the parent RFe$_2$Si$_2$ materials and on all Mn, Ni and Cu doped samples listed above. In all samples measured, no traces for SC were observed down to 5 K. On the other hand, the parent YFe$_2$Si$_2$ and LuFe$_2$Si$_2$ exhibit pronounce magnetic peaks at various temperatures in their magnetization M(T) curves. Generally speaking, the magnetic features of the Mn, Ni and Cu doped materials are very similar to that of their parent compounds. For the sake of clarity, we present first the extended data measured for YFe$_2$Si$_2$.

*(i) YFe$_2$Si$_2$* Fig. 3 shows the ZFC magnetization curve (M(T)) of YFe$_2$Si$_2$ measured at H= 250 Oe, in which the pronounced peak at 226 K is readily observed. The peak position depends strongly on the applied field. Fig. 3 (inset) also shows (*a*) that for H=1 kOe the peak shifts to 220 K and (*b*) that the same peak position is obtained regardless of whether the M(T) was measured via ZFC or FC processes. The bifurcation observed at low temperatures is probably due to the tiny ferromagnetic (FM) phase discussed hereafter. The field dependent of the peak position of YFe$_2$Si$_2$ is shown in Fig. 4. The peak obtained at 232 K for H=15 Oe, shifts to 118 K at 10 kOe (main panel) and further to 209, 197 and 186 K for, 20 30 and 45 kOe (inset) respectively. The almost linear field dependent of the peak position yields a slope of 1.2 K/kOe. It should be noted, that no observable peak was obtained in ac studies. The peak origin is discussed later.

| Compound | a(Å) ±0.001 | c (Å) ±0.003 | V (Å)$^3$ | Peak at 250 Oe (K) | $M_{Sat}$ ( 5 K) emu/g |
|---|---|---|---|---|---|
| **Y(Fe$_{1-x}$Ni$_x$)$_2$Si$_2$** | | | | | |
| x=0 | 3.923 | 9.951 | 153.1 | 226 | 0.31 |
| x=0.05 | 3.920 | 9.912 | 152.3 | 211 | 0.42 |
| x=0.15 | 3.923 | 9.874 | 151.9 | 192 | 0.13 |
| x=0.50 | 3.930 | 9.76 (1) | 150.7 | 54 | 1.82 |
| x=0.75 | 3.943 | 9.73(3) | 151.2 | 49 | 2.21 |
| x=1.0 | 3.963 | 9.544 | 149.8 | | |
| **Y(Fe$_{1-x}$Cu$_x$)$_2$Si$_2$** | | | | | |
| x=0.10 | 3.924 | 9.94 (1) | 153.0 | & | 0.008 |
| x=0.50* | 3.949 | 9.90 (2) | 154.4 | 170 | 0.014 |
| **Y(Fe$_{1-x}$Mn$_x$)$_2$Si$_2$** | | | | | |
| x=0.10 | 3.923 | 9.963 | 153.3 | 61 | |
| x=020 | 3.925 | 9.759 | 153.6 | 16 | |
| x=0.30 | 3.921 | 10.013 | 153.9 | | |
| x=0.50* | 3.921 | 10.129 | 155.7 | | |
| **Lu(Fe$_{1-x}$Ni$_x$)$_2$Si$_2$** | | | | | |
| x=0 | 3.875 | 9.869 | 148.2 | 70 | 0.01 |
| x=0.10 | 3.876 | 9.824 | 147.6 | 56 | |
| x=0.20 | 3.878 | 9.790 | 147.2 | 30 | 0.002 |
| x=0.30 | 3.881 | 9.761 | 147.0 | 23 | 0.013 |
| **La(Fe$_{1-x}$Ni$_x$)$_2$Si$_2$** | | | | | |
| x=0 | 4.059 | 10.163 | 167.4 | & | 0.96 |
| x=0.05 | 4.054 | 10.130 | 166.5 | & | 1.63 |
| x=0.15 | 4.054 | 10.10 | 166.0 | & | 2.34 |

\* contain extra lines (see text)    & peak> 350 K

Table 1. Lattice parameters, unit cell volume, the magnetic peak position at 250 Oe and the spontaneous magnetization at 5 K of R(Fe$_{1-x}$M$_x$)$_2$Si$_2$

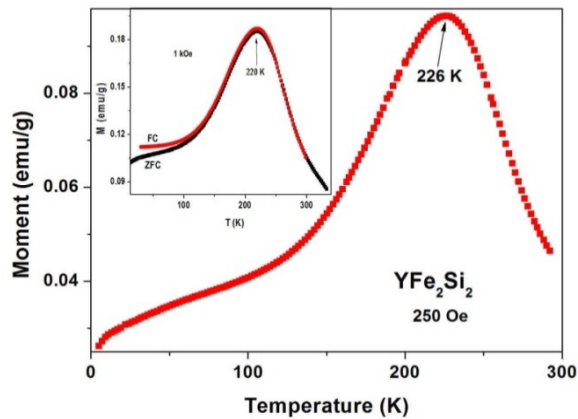

Fig. 3. (color online) ZFC magnetization curve of YFe$_2$Si$_2$ measured at 250 Oe. The inset shows the ZFC and FC branches measured at 1 kOe.

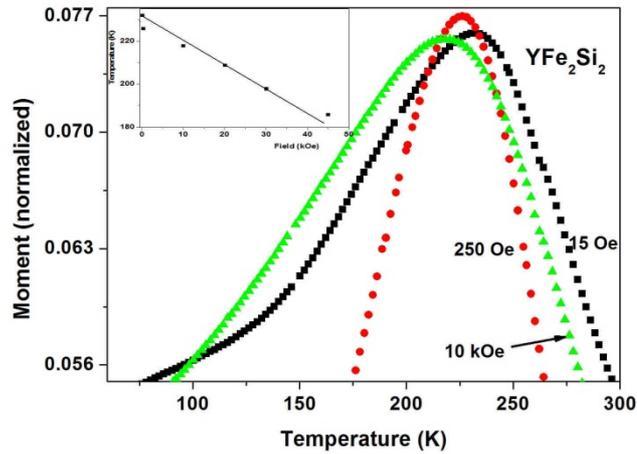

Fig. 4. (Color online) The field dependence of the peak position of YFe$_2$Si$_2$. The inset shows this behavior for high applied fields.

In order to clarify the nature of the peak in YFe$_2$Si$_2$, $^{57}$Fe MS below (90 K) and above (295 K) the peak position have been carried out. Fig. 5 shows two almost identical narrow singlet obtained with a line width of 0.333(3) mm/s. Due to the typical second order Doppler shift the two spectra differ only in their I.S. values, which are; 0.182±0.002 and 0.195±0.002 mm/s RT at 90 K respectively. Although the Fe ions in the 4d crystallographic site have a non-cubic symmetry, the quadrupole splitting, is relatively small, 0.06±0.02 mm/s. These values correspond to divalent nonmagnetic Fe state found in other Fe-based inter-metallic compounds with the ThCr$_2$Si$_2$-type structure [22, 23]. Fig. 5 definitely proves the absence of sizable permanent magnetic local moments in the Fe sites, and confirms our old claim that Fe in RFe$_2$Si$_2$ is non-magnetic [1-3].

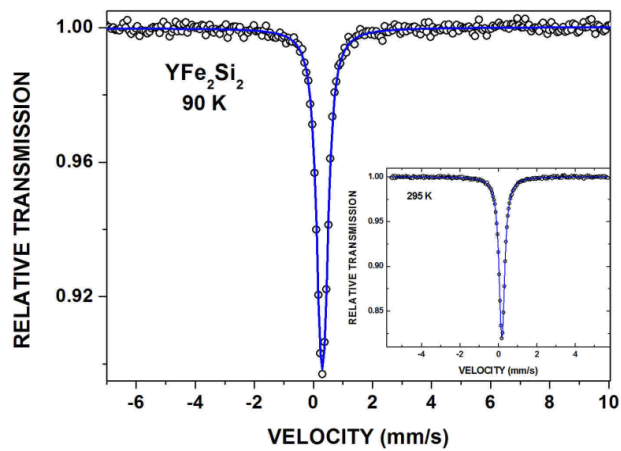

Fig. 5. (Color online) $^{57}$Fe Mössbauer studies of YFe$_2$Si$_2$ at 90 K (main panel) and at 295 K (inset)

The isothermal magnetization M(H) curves of YFe$_2$Si$_2$ measured at 5 and 295 K are shown in Fig. 6. These curves can be fitted as: M(H) = M$_S$ + $\chi_p$H, where M$_S$ is the spontaneous FM magnetization and $\chi_P H$ is the linear paramagnetic (PM) intrinsic susceptibility. For both temperatures, $M_S = 0.31(2)$ emu/g (Table 1) is attributed to a tiny FM extra phase with a very high magnetic temperature ordering which may exists in the sample such as: ~ 0.14% un-reacted pure Fe ($M_S = 220$ emu/g), $T_C=1040$ K), $Fe_xSi$ or $Fe_3Si$ ($T_C =808$ K).[24] The tiny amount of this FM impurity phase is below the detection limit of XRD and MS. The hysteresis loop at 5 K (shown in Fig. 6 (inset)) with a coercive field ($H_C$) of 245 Oe reflects this FM phase.

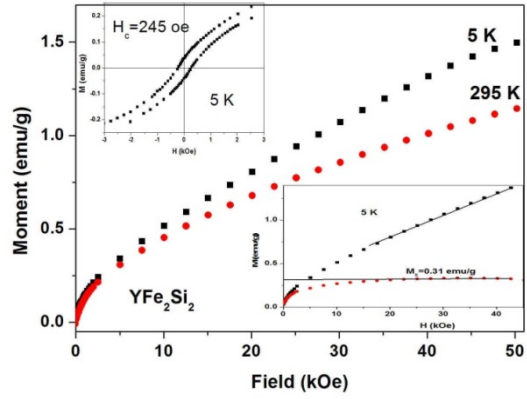

Fig. 6. (Color online) Isothermal magnetization M(H) plots measured at 5 and 295 K for $YFe_2Si_2$. The upper inset is the hysteresis loop measured at 5 K. The lower inset shows the two components deduced from the M(H) plot measured at 5 K. $M_S$ is the spontaneous FM magnetization of the impurity (Table 1) and $\chi_p H$ is the PM susceptibility.

Well above the peak position the M(T) curves exhibit a typical PM shape and adhere closely to the Curie-Weiss (CW) law: $\chi(T) = \chi_0 + C/(T-\theta)$, where $\chi(=M/H)$, $\chi_0$ is the temperature independent part $C$ is the Curie constant, and $\theta$ is the CW temperature. The PM parameters extracted from M(T) measured at 1 kOe are: $\chi_0 =5.5\cdot10^{-3}$ emu/mol Oe, $C = 2.34(1)$ emu K/mol Oe and $\theta = 191(3)$ K. The high $\chi_0$ is inferred to the temperature independence FM impurity as discussed. This C value corresponds to a PM effective moment of $P_{eff}= 3.06$ $\mu_B$/ Fe. $\chi(T)$ obtained fits well with $\chi_p$ deduced at 5 K. The $\theta$ value obtained, reminiscences an anti-ferromagnetic structure for $YFe_2Si_2$, as reported in Ref. 25. However, our Mössbauer studies (Fig. 5) definitely exclude this possibility. It appears that the Fe in $YFe_2Si_2$ carries a net PM moment. That is in stark contrast to virtually temperature-independent magnetic susceptibility (Pauli PM) claimed in the past.[1] Note, that the peak size magnitude for $YFe_2Si_2$ (Fig. 3) is ~$10^{-2}$ emu/g, which is below the old magnetometer sensitivity used in Ref. 1.

*(ii) $LaFe_2Si_2$ and $LuFe_2Si_2$.* The ZFC magnetic behavior of $LaFe_2Si_2$ is a bit different. In contrast to $YFe_2Si_2$ (Fig. 4), the magnetization measured at H=250 Oe increases up to 300 K but no definite peak is observed (Fig. 7). Presumably, higher temperatures are needed to reveal this peak as observed for the iso-structural $LaNi_2Ge_2$ material [26]. A small bump around 225 K is observed, its nature is yet not known. On the other hand, for $LuFe_2Si_2$ the peak at 70 K under H=250 Oe is readily observed. The isothermal magnetization curves at 5 K of $LaFe_2Si_2$ and $LuFe_2Si_2$ (Fig. 7 lower inset) can be fitted in the same manner as the M(H) plots of $YFe_2Si_2$ (Fig. 6). The deduced FM $M_S$ values are: 0.96 and 0.01 emu/g respectively (Table 1). This explains as to why the measured moment values (~ 0.002 emu/g, Fig.7) for $LuFe_2Si_2$ are much lower than these of $YFe_2Si_2$ and $LaFe_2Si_2$. The almost straight line at 5 K for $LuFe_2Si_2$ excludes the existence of the SC $Lu_3Fe_3Si_5$ phase ($T_C= 6$ K).[27] The main goal in presenting the M(T) curves for the three $RFe_2Si_2$ samples (Figs. 3, and 7), is to show the appearance of pronounced magnetic peaks and their locations. Therefore, no subtraction of the impurity FM phases contribution was made. Since in all $RFe_2Si_2$ (R=La, Y and Lu) R are not magnetic elements, the only difference between the three samples is in their unit-cell dimensions.

Table 1 shows that the lattice parameters as well as the unit cell volumes decrease as: LaFe$_2$Si$_2$ > YFe$_2$Si$_2$ > LuFe$_2$Si$_2$. The same trend is observed in peak position (measured at the same H), indicating clearly the connection between the lattice dimensions and the peak positions.

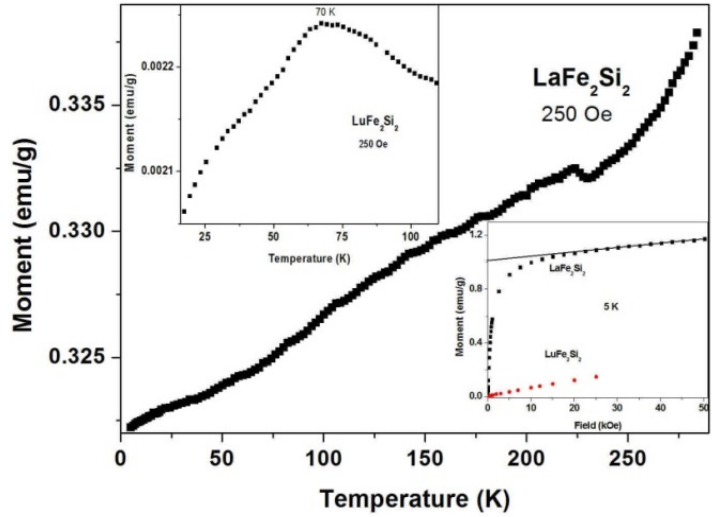

Fig. 7 ZFC magnetization curve of LaFe$_2$Si$_2$ and LuFe$_2$Si$_2$ (upper inset) measured at 250 Oe. The lower inset shows the isothermal M(H) plots of LaFe$_2$Si$_2$ and LuFe$_2$Si$_2$ measured at 5 K. Note the almost linear curve of LuFe$_2$Si$_2$.

*(c)* **Magnetic studies of R(Fe$_{1-x}$Ni$_x$)$_2$Si$_2$**

As stated above, the magnetic behaviors of all R(Fe$_{1-x}$Ni$_x$)$_2$Si$_2$ samples resemble the features observed in their parent compounds and the relevant data obtained are listed in Table 1. For each x value, the M(T) curves show distinct peaks at elevated temperatures which also decreases with H. All M(H) plots contain an extra FM phase similar to that shown in Figs. 6-7. The differences among the samples are (i) the peak position which is strongly depends on x, (ii) the spread M$_S$ values as listed in Table 1.

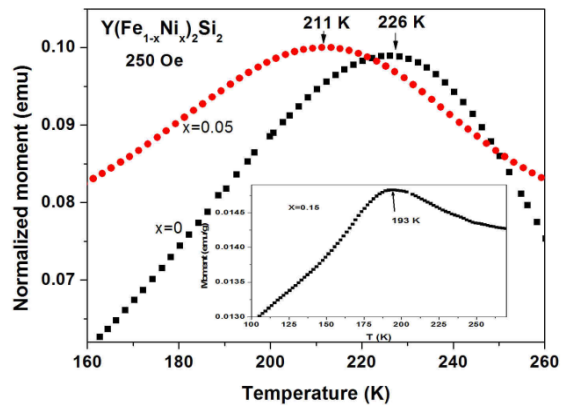

Fig. 8. (Color online) ZFC magnetization curve of Y(Fe$_{1-x}$Ni$_x$)$_2$Si$_2$ (x=0, 0.05 0.15 (inset)) measured at 250 Oe.

*Y(Fe$_{1-x}$Ni$_x$)$_2$Si$_2$*. (i) Figs. 8-10 show representative M(T) plots of Y(Fe$_{1-x}$Ni$_x$)$_2$Si$_2$ samples, all measured at 250 Oe. Fig. 8 shows the M(T) for x=0, 0.05 and 0.15. The peak at 226 K for YFe$_2$Si$_2$ (x=0) shifts to 211 and 192 K for x=0.05 and 0.15 and further down to 54 and 50 K for x=0.5 and 0.75 respectively.

Note, that the peak is visible even at this low Fe concentration (Fig. 10). The decrease of the peak position with x measured at lower applied fields (15-25 Oe) is shown in Fig. 10 (inset). (ii) For each Y(Fe$_{1-x}$Ni$_x$)$_2$Si$_2$ material, the peak is also field depended. For example: (a) for x=0.05, the peak at 220 K under H=15 Oe shifts to 211 K and to 193 K when measured at 250 and 1 kOe respectively and (b) for x=0.15, the peak shown in Fig 8, shifts to 180 K when measured at 1 kOe. (iii) All M(H) plots measured below and above the peak position are similar to these YFe$_2$Si$_2$ (Fig. 6) and can be fitted by the same manner. (see Fig. 9 inset). M$_S$ deduced at 5 K ranged from the lowest 0.13 emu/g for x=0.15 to 2.21 emu/g deduced for x=0.75 (Table 1). This highest M$_S$ value corresponds to ~1 % Fe as an impurity phase, a level which is undetectable neither by XRD nor by MS.

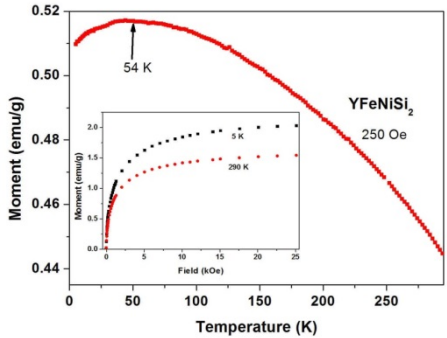

**Fig. 9.** (Color online) M(T) of Y(Fe$_{0.5}$Ni$_{0.5}$)$_2$Si$_2$ measured at 250 Oe. The inset shows the isothermal M(H) at 5 K and 290 K. Note the similarity to the M(H) curves shown in Fig. 6.

An additional point of interest is the magnetic behavior of YNi$_2$Si$_2$ (Fig. 11) which is presented here for the first time. The lattice parameters obtained for YNi$_2$Si$_2$ (Table 1) are in fair agreement with data published in Ref 20. The M(T) curve depends on the temperature (not a Pauli behavior) but all attempts to fit it to the CW law were unsuccessful. Both M(H) curve measured at 5 K and at 250 K are linear and the calculated $\chi$(=M/H) value at 5 and 246 K are 3.76·10$^{-4}$ and 1.96·10$^{-4}$ emu/mol Oe respectively. This linear behavior strengthens our suggestion that the undetermined extra FM phase is due to Fe or to any other Fe intermetallic compound.

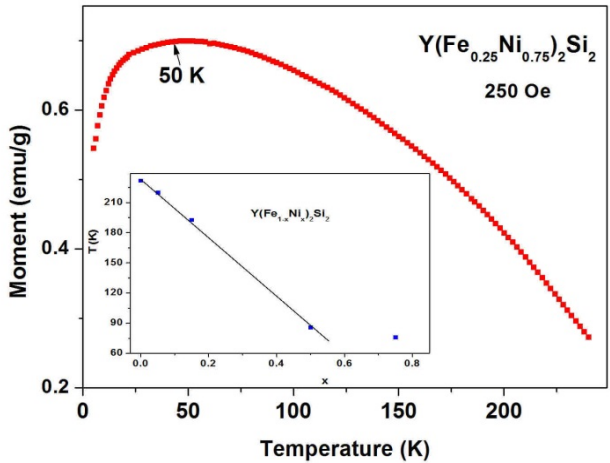

**Fig. 10.** (Color online) Temperature dependence of the magnetization of Y(Fe$_{0.25}$Ni$_{0.75}$)$_2$Si$_2$ measured at 25 Oe. The inset shows the peak position of all Y(Fe$_{1-x}$Ni$_x$)$_2$Si$_2$ samples measured at 20-30 Oe.

*La(Fe$_{1-x}$Ni$_x$)$_2$Si$_2$ and Lu(Fe$_{1-x}$Ni$_x$)$_2$Si$_2$*. In these two systems only materials with low Ni concentration have been synthesized. Their M(T) and M(H) plots are very similar to their parent compounds. The relevant data obtained are listed in Table 1. Although the peak position is strongly depends on x, no peak is observed in La(Fe$_{0.85}$Ni$_{0.15}$)$_2$Si$_2$ up to 385 K. Here again a bump in the M(T) is observed around 260 K (Fig. 12). The $^{57}$Fe MS measured at 90 K exhibits a narrow singlet with almost the same parameters obtained for YFe$_2$Si$_2$. This proves once again the absence of sizable permanent magnetic local moments in the Fe sites. For La(Fe$_{1-x}$Ni$_x$)$_2$Si$_2$, the deduced M$_S$ values at 5 K increases with x.

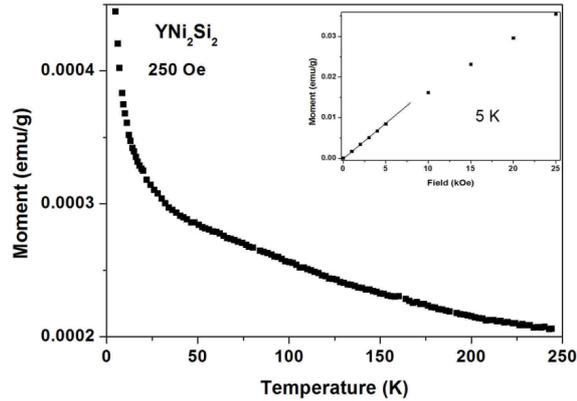

Fig. 11. Temperature dependence of the magnetization of YNi$_2$Si$_2$ measured at 250 Oe. The inset shows the almost linear isothermal magnetization at 5 K.

Similar to Y(Fe$_{1-x}$Ni$_x$)$_2$Si$_2$, in the *Lu(Fe$_{1-x}$Ni$_x$)$_2$Si$_2$* system the peak position shifts with x to lower temperatures (Fig. 13). At 250 Oe, the peaks for x=0.1 and 0.2 are at 56, 30 K respectively (Table 1). For x=0.2, the peak is shifted to 43 K when measured at 25 Oe. For both materials, an additional broad peak is observed at 205 K (not shown) its origin is yet to be determined. For x=0.3, at 100 Oe the peak location is at 31 K (Fig. 13) and it is shifted to 18 K when measured at 1 kOe. The high temperature range of the M(T) curve for x=0.3 also exhibit a typical PM shape and can be fitted by the CW law. The extracted PM parameters (1 kOe) are: $\chi_0$ =4.2·10$^{-3}$ emu/mol Oe, C = 2.60(2) emu K/mol Oe and $\theta$ = -70(1) K. This C value corresponds to P$_{eff}$= 3.71 $\mu_B$/ Fe and proves once more the PM nature of Fe in the RFe$_2$Si$_2$ systems. The M(H) curves of *Lu(Fe$_{1-x}$Ni$_x$)$_2$Si$_2$* are almost linear and the deduced M$_S$ values are small and negligible (Table 1).

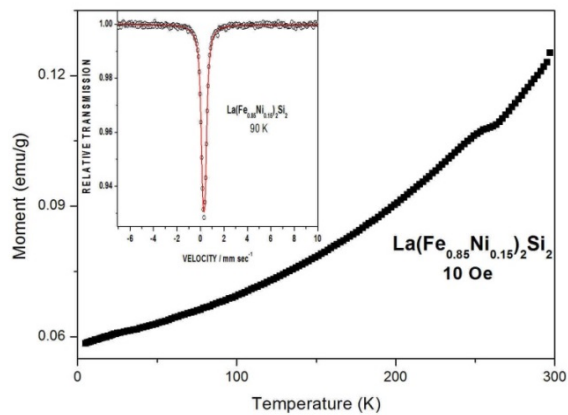

Fig. 12. (Color online) Temperature dependence of the magnetization of La(Fe$_{0.85}$Ni$_{0.15}$)$_2$Si$_2$ measured at 10 Oe. The inset shows the $^{57}$Fe Mössbauer spectrum at 90 K.

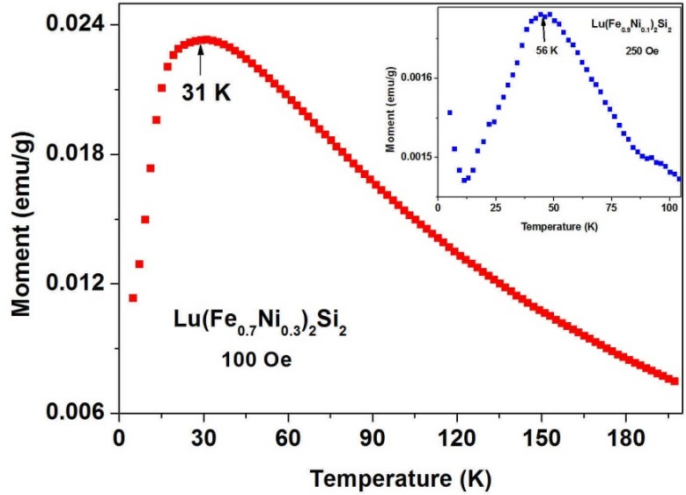

Fig. 13. (Color online) Temperature dependence of the magnetization of Lu(Fe$_{0.7}$Ni$_{0.3}$)$_2$Si$_2$ and Lu(Fe$_{0.9}$Ni$_{0.1}$)$_2$Si$_2$ (inset) measured at 100 Oe and 250 Oe respectively.

(d) **Magnetic studies of the Y(Fe$_{1-x}$Cu$_x$)$_2$Si$_2$ system**

The change in lattice parameters Y(Fe$_{1-x}$Cu$_x$)$_2$Si$_2$ is a bit different than that of R(Fe$_{1-x}$Ni$_x$)$_2$Si$_2$ systems. Table 1 shows that for Cu the *a* lattice parameter increases with x whereas the *c* constant practically remains unchanged. This agrees well with the higher *a* lattice constant of YCu$_2$Si$_2$ (a=4.153 and c=9.92 Å) as compared to that of YFe$_2$Si$_2$.[20] In the ThCr$_2$Si$_2$ type structure the shortest 3d-3d distance is given by given by $2^{-1/2}a$, which means that the Cu size is larger than that of Fe. That is due to the well accepted determination that in many intermetallic compounds Cu is formally monovalent, in contrast to a formal divalent Fe (and also Ni) deduced from MS spectra (Fig. 5). Thus, the Fe-Cu bonds are larger than Fe-Fe and/or Fe-Ni bonds. That probably affects the peak position of Y(Fe$_{1-x}$Cu$_x$)$_2$Si$_2$. Indeed, Fig.14 shows that for x=0.1, the M(T) measured at 250 Oe, increases up to 300 K but no definite peak is observed. At 5 kOe, a shallow peak is obtained at 302(2) K. In addition, a sharp upeak is observed at 24 K (not shown) its nature is not yet known.

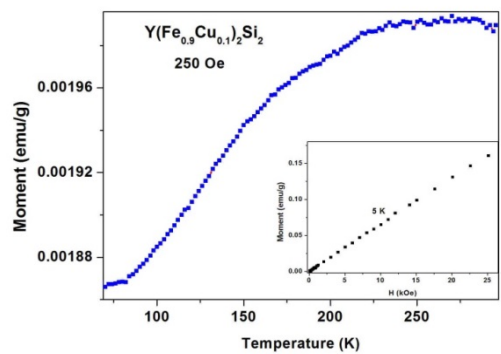

**Fig. 14.** (Color online) Temperature dependence of the magnetization of $Y(Fe_{0.9}Cu_{0.1})_2Si_2$ measured at 250 Oe. The inset shows the linear isothermal magnetization at 5 K.

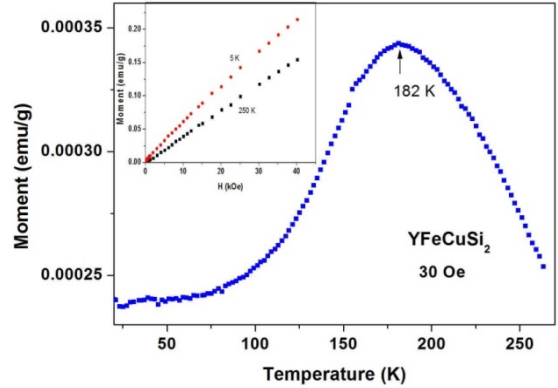

**Fig. 15.** (Color online) Temperature dependence of the magnetization of $Y(Fe_{0.5}Cu_{0.5})_2Si_2$ measured at 30 Oe. The inset shows the linear isothermal magnetizations at 5 K and 250 K which are below and above the peak position.

As expected, increase of Cu concentration shifts the peak position to lower temperatures, as shown in Fig. 15 for $YFeCuSi_2$ (x=0.5). The peak at 182 K (at H=30 Oe) shifts to 170 and to 166 K for H=250 Oe and 1 kOe respectively. Thus, the applied magnetic field affects the $Y(Fe_{1-x}Cu_x)_2Si_2$ system in the same manner. All M(H) curves are almost linear (Figs. 14-15 (insets)) and the FM impurities are negligible. The hexagonal YCuSi extra phase detected by XRD is not magnetic and cannot account for peaks observed in Fig.15. [28]

(e) **Magnetic studies of the $Y(Fe_{1-x}Mn_x)_2Si_2$ system**

The $RMn_2Si_2$ system was extensively studied in the past. All compounds are magnetically ordered at high temperatures and the interactions between the Mn-Mn moments within the layers are always ferromagnetic. The nature of the interlay exchange interaction is FM or AFM, depending on the value of the lattice constants. e.g. $LaMn_2Si_2$ is FM with $T_C$=310 K whereas $YMn_2Si_2$ is AFM with $T_N$=510 K.[29] Using the $^{57}Fe$ MS we have shown that (i) Fe in the doped materials has no magnetic moment of its own and it reveals the magnetic order of the Mn sublattice through transferred hyperfine field, (ii) that above 310 K, $LaMn_2Si_2$ becomes AFM ordered up to 470 K. [30]. (iii) In contrast to the systems described above Mn in $Y(Fe_{1-x}Mn_x)_2Si_2$ carries a local magnetic moment. [23] The major goal here is study the effect of magnetic Mn on peak positions in $YFe_2Si_2$.

XRD studies on $Y(Fe_{1-x}Mn_x)_2Si_2$ (x=0.1, 0.2, 0.3,0.5, 0.7 and 1) show that single phase materials could be obtained for the low Mn compositions only. In $YFeMnSi_2$ (x=0.5) a few extra lines which belong to the orthorhombic $YFe_{0.33}Si_2$ have been observed. In addition, for x=0.7 and $YMn_2Si_2$ (x=1) extra lines which were indexed as belong to the AFM $Y_2Mn_3Si_5$ ($T_N$=96 K) phase were detected [31]. Therefore, their magnetic properties will not be discussed here.

The effect of magnetic Mn substitution on the peak position is more dramatic. Fig. 16 (inset) exhibits the M(T) curve for $Y(Fe_{0.9}Mn_{0.1})_2Si_2$ measured at 250 Oe in which the peak at 63 K is well observed. Although, the lattice parameters of this material and pure $YFe_2Si_2$ are very similar (Table 1), the shift in the peak location is ~160 K as compared to ~40 K only in $Y(Fe_{0.85}Ni_{0.15})$. Further increase of x to 0.2, shifts the peak to 16 K at 250 Oe and to 13 K at 1 kOe (Fig. 16). No peak is observed for the x=0.3 sample. Fig. 16 also shows some decrease in the M(T) curve around 160 K which reminiscences a FM like behavior with a transition ($T_M$) ~ 220 K, which is discussed hereby. $^{57}$Fe MS spectra taken at 92 and 295 K (not shown) are identical, indicating an absence of any local magnetic field on Fe nuclei.

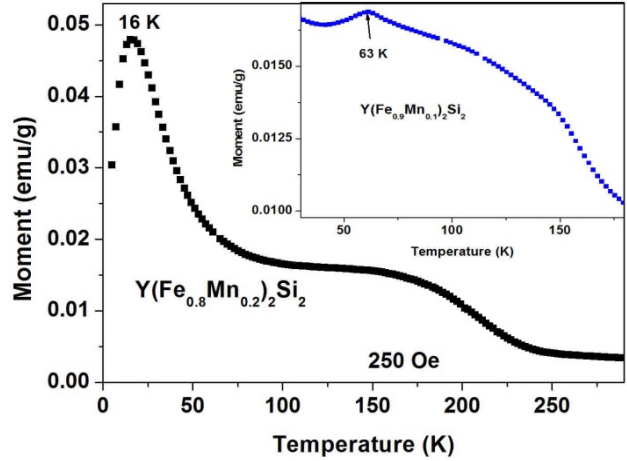

Fig. 16. (Color online) Temperature dependence of the magnetization of Y(Fe$_{0.8}$Mn$_{0.2}$)$_2$Si$_2$ and Y(Fe$_{0.9}$Mn$_{0.1}$)$_2$ (inset) both measured at 250 Oe

Different features are observed in YFeMnSi$_2$. Fig. 17 shows a sharp decrease in M(T) curve up to 29 K and then a rise up to ~ 100 K and the FM like behavior with $T_M$ ~250 K. Note, the increase of $T_M$ with x. $^{57}$Fe MS spectra taken at 90 K and RT indicate clearly that the 90 K spectrum is much broader than the RT one. Least square fits of the two spectra yield the hyperfine parameters as follows: The line width in both spectra is 0.33(3) mm/s. At RT the small quadrupole splitting is 0.12±0.02 mm/s and the I.S is 0.205±0.002 mm/s, values which agree well with those obtained for YFe$_2$Si$_2$ presented above. The broader line at 90 K (I.S. =0.30 mm/s) was fitted by adding a magnetic hyperfine field parameter $H_{eff}$ =30.3(3) kOe. This $H_{eff}$ value is an order of magnitude small than $H_{eff}$ =330 kOe for Fe. That indicates that at 90 K, Fe in YFeMnSi$_2$ senses transferred magnetic fields produced by the magnetic Mn ions. The sharp increase of the bulk magnetization below 29 K, is probably attributed to reorientation of the Mn magnetic sublattice a phenomenon which is very common to the RMn$_2$Si$_2$ system.[32] A supporting evidence for this reorientation is given by the M(H) plots shown in Fig. 17 (inset). Above the magnetic transition (at 290 K) a linear M(H) curve is obtained. On the other hand, the two nonlinear M(H) plots at 5 and 90 K, overlap at H> 25 kOe, but exhibit a quite different shape at low H values. At 5 K only, a small hysteresis loop is developed with a coercive field of 235 Oe. [For Y(Fe$_{0.3}$Mn$_{0.7}$)$_2$Si$_2$ (x=0.7) the $^{57}$Fe MS spectrum at RT is also broad and is fitted with $H_{eff}$= 39.2 kOe, indicating $T_M$ >RT. Note that $H_{eff}$ increases with x].

Using the same token, we may analyze the magnetic curve for Y(Fe$_{0.8}$Mn$_{0.2}$)$_2$Si$_2$ as composed of two magnetic contributions: the peak at 16 K arises definitely from the Fe ions and (ii) the FM like behavior stems from the Mn ions. (For nominal x=0.2, the Mn concentration is close to the two dimensional percolation limit). Due to the huge shift in the peak position caused by the magnetic Mn ions, we tend to believe that these two components coexist. If that is the case, the produced transferred hyperfine field is very small and is below the sensitivity of the Mössbauer technique. On the other hand, a phase separation case cannot be excluded. For this scenario, the absence of $H_{eff}$ on the Fe nuclei is obvious.

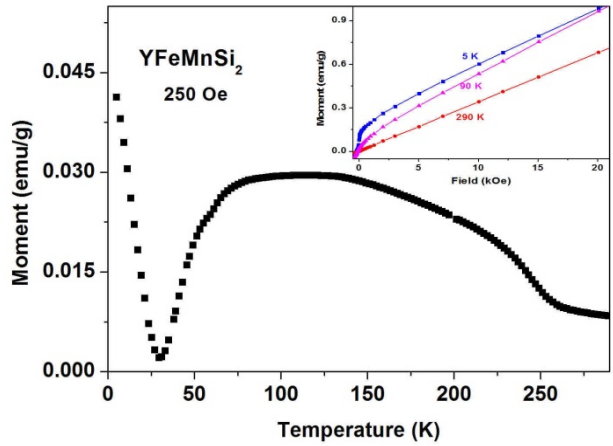

Fig. 17. (Color online) Temperature dependence of the magnetization of YFeMnSi$_2$ measured at 250 Oe. The inset shows three isothermal magnetization plots measured at 5, 90 and 290 K.

We are turning now to the small deduced H$_{eff}$ =30.3 kOe in YFeMnSi$_2$ at 90 K. If the Fe had its own magnetic moment, then H$_{eff}$ should not be so sensitive to x and rather will be in full saturation with H$_{eff}$ of an order of magnitude higher. But if the H$_{eff}$ is transferred from neighboring magnetic Mn ions, then each Fe which has other Mn ions as first neighbors will experience strongly reduced hyperfine fields. Assuming random distribution of Mn in the 3d layers, 94% of the Fe nuclei (given by $(1-(1-x)^4)$) should experience reduced fields which lead to the broadening of the spectrum at 90 K. Further calculations of H$_{eff}$ for other x values, is beyond the scope of this paper. However the H$_{eff}$=30.3 kOe obtained, fits well with the calculated H$_{eff}$ values for the Y(Fe$_{1-x}$Mn$_x$)$_2$Si$_2$ system and to measured values on similar systems. [30, 32-33].

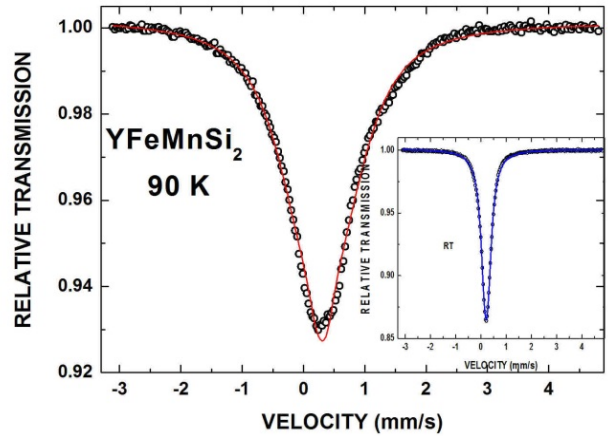

Fig. 18. (Color online) $^{57}$Fe Mössbauer studies of YFeMnSi$_2$ at 90 K (main panel) and at 295 K (inset)

**(iv) Discussion and conclusions**

Single or nearly single-phase polycrystalline samples of R(Fe$_{1-x}$Ni$_x$)$_2$Si$_2$ (R=La, Y and Lu), Y(Fe$_{1-x}$Cu$_x$)$_2$Si$_2$ and Y(Fe$_{1-x}$Mn$_x$)$_2$Si$_2$ were synthesized and their magnetic properties have been measured. XRD patterns showed that the samples crystallize in the BCT ThCr$_2$Si$_2$ type structure. Because of the smaller ionic radius of Ni, the unit cell volume in all R(Fe$_{1-x}$Ni$_x$)$_2$Si$_2$ systems, decrease with x. On the other hand, for Y(Fe$_{1-x}$Mn$_x$)$_2$Si$_2$ and Y(Fe$_{1-x}$Cu$_x$)$_2$Si$_2$ the unit cell volume increases with x. This is caused by the larger Mn radius and by the formal monovalent state of Cu in intermetallic compounds.

In both HTSC cuprates and Fe-As based materials the SC phase evolves from an AFM (or SDW) parent compound and it was proposed that the magnetic state is essential to mediate SC. [4,34]. As stated above, the similarity between the two $BaFe_2As_2$ and $RFe_2Si_2$ systems motivated us to look for SC in various doped $RFe_2Si_2$ materials, although Fe is not magnetic. None of the samples listed above is SC above 5 K. This piece of evidence may shed some light on the SC mechanism, by proving the necessity of magnetic interactions in the parent compounds in order to form the SC state. However, this assumption is not conclusive, because the full phase-diagram of the R-Fe-Si is not yet known. One may argue that magnetic ordering might exist in compositions other than 1:2:2, thus all the magnetic part in the full phase diagram is missing, therefore the absence of magnetic ordering in the stoichiometric 1:2:2 materials is just accidental.

Two characteristic features in the dc magnetization studies are observed. (a) Pronounced peaks in the M(T) plots for all systems (except for R=La). Four independent factors may affect the peak position from which the first and the last factors are more drastic than the rest. (i) It is very sensitive to the lattice parameters. Increasing and/or decreasing of the unit cell volume change this temperature accordingly. In the parent $RFe_2Si_2$ materials, for R=La the peak is well above RT, but for R=Y and Lu (measured at 250 Oe) the peak is at 226 K and 70 K respectively. In the doped $R(Fe_{1-x}M_x)_2Si_2$ samples, either (ii) increasing of x, (iii) applying magnetic fields or (iv) magnetic ions (such as Mn) all shift the peak to lower temperatures. The peak at 226 K for $YFe_2Si_2$ is lowered by ~40 or ~160 K in $Y(Fe_{0.85}Ni_{0.15})_2Si_2$ and $Y(Fe_{0.9}Mn_{0.1})_2Si_2$ respectively and absences in $Y(Fe_{0.7}Mn_{0.3})_2Si_2$.

No peak is observed in $YNi_2Si_2$ whose magnetic features are first reported here. Therefore, we attribute all peaks observed to the Fe sublattice. The absence of magnetic hyperfine field on the Fe nuclei (Fig. 5) in $YFe_2Si_2$ (and in other compounds) confirms that Fe has no sizable local moment by its own. In $Y(Fe_{1-x}Mn_x)_2Si_2$ the small transferred hyperfine field observed on Fe, is induced by the magnetic Mn sublattice

(b) A tiny extra FM phase with very high $T_C$, is obtained in all $Y(Fe_{1-x}Ni_x)_2Si_2$ samples whereas in the rest $Y(Fe_{1-x}M_x)_2Si_2$ (M=Cu or Mn) materials the impurity phase is negligible. The un-avoidable impurity phases probably due to unreacted Fe (or from $Fe_xSi_y$) contribute constant moment to the various M(T) plots and change their absolute values, but the do not affect the magnetic features of the bulk materials. Thus the main issue remains is the nature of the peaks observed in the systems studied here, which presumably all have the same origin. For the sake of simplicity we'll discuss the magnetic state of $YFe_2Si_2$.

The shortest Fe-Fe distance (given by $2^{-1/2}a$) in $YFe_2Si_2$, is 2.77Å. This distance is much longer than the critical value of 2.52 Å for FM interactions in Fe based intermetallic compounds. [35] Thus no long range magnetic order exists in this sample at any temperature, as determined by our MS studies (Fig. 5). $YFe_2Si_2$ was claimed to be AFM ordered [22] or FM ordered at $T_C$= 790 K. [36]. Both determinations are shown to be wrong, with the high reported $T_C$, being just that of the impurity phase (probably $Fe_3Si$). In previous publications, it was proposed that for non-magnetic R in $RFe_2Si_2$, the divalent Fe ions are Pauli paramagnet with a virtually temperature-independent magnetic susceptibility. [1,37] The present measurements using the high sensitive SQUID magnetometer reveals that Fe in $YFe_2Si_2$ and in all its derivatives is rather paramagnetic. This is supported by (1) the high temperature M(T) region which follows the CW law from which an average effective moment of $P_{eff} \sim 3.4$ $\mu_B$/ Fe is deduced; (2) from all isothermal M(H) curves in which the PM component is very pronounced.

Maximum in the susceptibility (around 70 K) was already observed for pure Pd metal which is considered as the classic example of nearly ferromagnetic Fermi liquid (NFFL), with large effect of spin fluctuations. With its high DOS at the Fermi energy and large Stoner exchange enhancement (~ 10), Pd is easily polarized by dilute magnetic moments and often to FM at relatively high temperatures.[38] This maximum is a common feature in other NFFL compounds such $YCo_2$, and $LuCo_2$ [39]. Quantitative calculation of the susceptibility is still a challenge even for the simplest case of Pd.

$YFe_2Zn_{20}$ (with Stoner enhancement factor of $Z=0.88$) is closer to FM than Pd [40-41] and also serves as an archetypical examples of NFFL. A faint maximum in the susceptibility around 10 K was obtained when measured at low magnetic fields. This maximum is monotonically decreased with increasing H. Another example is the iso-structural $LaNi_2Ge_2$ in which a peak in the susceptibility around 400 K was observed. The peaks origin is not clear yet, and it is attribute it to low-dimensional AFM correlations.[26] A shallow peak is also observed in $SrCu_2As_2$ but its appearance is ignored and not discussed. [42]

We tend to believe that $YFe_2Si_2$ and its derivatives studied here are additional examples for NFFL systems. In contrast to all examples mentioned, here either the lattice parameters and/or the applied field, has a pronounced effect on the peak position. Doping of non-magnetic Ni ions also shift the peak to lower temperatures, but this shift is more dramatic when non-zero magnetics ions such as Mn are introduced. The current state of experiments does not allow us to suggest any consistent explanation to the phenomena presented here, which need more consideration. Whatever the explanation is, the discovery of a pronounced peak in $YFe_2Si_2$ is challenging for the theory of magnetism and propitious to further theoretical and experimental investigations. The study of these systems is a topic of ongoing research and promises to be a fruitful new phase space for several years to come.

**Acknowledgments**: The research in Jerusalem was partially supported by the joint German-Israeli DIP project and by the Klachky Foundation for Superconductivity and in Houston by US AFOSR and TCSUH. We thank Prof. I. Nowik carrying out the Mössbauer studies.